\documentclass[12pt,preprint]{aastex}

\begin{document}

\title {A TIDALLY-DISRUPTED ASTEROID AROUND THE WHITE DWARF G29-38}

\author{M. Jura} 
\affil{Department of Physics and Astronomy, University of California,
    Los Angeles CA 90095-1562; jura@clotho.astro.ucla.edu}

\begin{abstract}

The infrared excess around the white dwarf G29-38 can be explained
by emission from an opaque flat ring of dust with an inner radius of 0.14 R$_{\odot}$ and an outer radius $<$ 1R$_{\odot}$.  This ring lies within
the Roche region of the white dwarf where an  asteroid could have been tidally destroyed, producing a system reminiscent of Saturn's rings.   Accretion onto the white dwarf from this circumstellar dust can
explain the observed calcium abundance in the atmosphere of G29-38. Either as a bombardment by a series of asteroids or because of one large disruption, the total amount of matter  accreted onto the white dwarf may have been ${\sim}$ 4 ${\times}$ 10$^{24}$ g, comparable to the total mass of asteroids in the Solar System, or, equivalently, about 1\% of the mass in the asteroid belt around
the main sequence star ${\zeta}$ Lep.        

\end{abstract}

\section{INTRODUCTION}
The formation and evolution of large rocky solids such as  asteroids and terrestrial planets  are of great interest.  Even though ever-increasing numbers of extra Solar System Jovian-mass gas-giant planets  are  known (see Marcy \& Butler 1998), no rocky planet orbiting a main sequence star outside of our own Solar System has been directly discovered. There
is good indirect evidence for an asteroid belt around the nearby (distance = 22 pc) main sequence A-type star, ${\zeta}$ Lep  
(Chen \& Jura 2001) and there are rocky planets  around the pulsar
PSR 1257+12 (Wolszczan \& Frail 1992).   
Here, we propose that the infrared excess around the white
dwarf G29-38  is the consequence of
the tidal disruption of an asteroid.
  
Most white dwarfs have infrared colors characteristic of their photospheres,
but  G29-38 was discovered by Zuckerman \& Becklin (1987) to have
an infrared excess which can be explained by circumstellar dust at a temperature of approximately 1000 K.  The
origin of these warm grains which must be close to the white dwarf is an unsolved mystery (Zuckerman 2001).

 Graham et al. (1990) proposed that the circumstellar dust around G29-38
arose from an unspecified catastrophic destruction of an asteroid or comet near the white dwarf.   
Here, we present a model where the dust around G29-38 is so close to the star that it lies inside the Roche radius where an asteroid parent-body  would have been tidally disrupted,  a process  suggested by Debes \& Sigurdsson (2002) to
destroy comets and that would produce the observed circumstellar dust.  
Our model, where a flat, opaque dust ring ultimately evolves from a tidally-disrupted asteroid, may solve another mystery: the presence of calcium and other metals in the atmosphere of G29-38 (see Koester, Provencal \& Shipman 1997, Zuckerman \& Reid 1998) where the calculated dwell time for this material is only ${\sim}$ 20 years (Paquette et al. 1986).      Alcock, Fristrom \& Siegelman (1986) suggested that the
atmospheric composition of a white dwarf could be detectably altered by
the direct impact of a comet.  In the specific case of G29-38, we suggest a somewhat similar scenario where enough accretion to pollute the atmosphere
arises from the tidal disruption of an asteroid.  The model that accretion comes from the disruption of asteroids or comets  naturally explains the absence
of much hydrogen in the infalling material -- a major difficulty with the commonly suggested hypothesis (Dupuis, Fontaine \& Wesemael 1993) that white dwarf atmospheres are contaminated by accretion of interstellar matter.    

\section{MODEL FOR THE DUST DISTRIBUTION AND EMISSION}

In order to model the circumstellar dust emission, we need to characterize
G 29-38, a ZZ Cet variable (see Kleinman et al. 1998).  We adopt a distance of 14.1 pc (Tokunaga, Becklin \& Zuckerman 1990), a stellar radius, $R_{*}$, of 8.2 ${\times}$ 10$^{8}$ cm, an effective temperature, $T_{*}$, of 11820 K, and a  mass, $M_{*}$, of 0.69 M$_{\odot}$ (Bergeron et al. 1995).  These parameters yield a mean density, ${\rho}_{*}$, of 6.0 ${\times}$ 10$^{6}$ g cm$^{-3}$,  a stellar luminosity of 2.4 ${\times}$ 10$^{-3}$ L$_{\odot}$ and imply a cooling age  of
4 ${\times}$ 10$^{8}$ yr (Winget et al. 1987).  

We assume that this star illuminates a passive, opaque, flat circumstellar ring where the
incident optical energy is re-radiated in the infrared.   When
variations in the vertical ring structure are ignored, the standard solution  for the 
temperature, $T_{ring}$, as a function of
distance from the star, $R$, is given by the expression (Chiang \& Goldreich 1997):
\begin{equation}
T_{ring}\;{\approx}\;\left(\frac{2}{3{\pi}}\right)^{1/4}\,\left(\frac{R_{*}}{R}\right)^{3/4}\;T_{*}
\end{equation}
The predicted temperature of the ring from
equation (1) is lower than that expected for  unshielded, isolated black bodies, and the very innermost grains are likely to be warmer than predicted by equation (1).  As with Saturn's rings, the vertical thickness of the ring around G29-38 may not be much
greater than the size of the  particles (Nicholson 2000), which  typically is less than
10 cm (Graham et al. 1990).  Thus the  amount of unshielded material compared to the total mass in the ring may
be tiny.

The predicted flux from the ring, $F_{ring}$ is:
\begin{equation}
F_{ring}\;=\;\frac{2\,{\pi}\,cos\,i}{D_{*}^{2}}\;{\int}_{R_{in}}^{R_{out}}\;B_{\nu}(T_{ring})\,R\,dR
\end{equation}
where $D_{*}$ is the distance to the star from the Sun, $R_{in}$ and $R_{out}$ represent the inner and outer radii of the dust
ring, $i$ is the inclination angle of the ring and $B_{\nu}$ is the Planck function.
 With the substitution that 
$x$ = $h{\nu}/kT_{ring}$, then:
\begin{equation}
F_{ring}\;{\approx}\;12\,{\pi}^{1/3}\,\left(\frac{R_{*}^{2}\,cos\,i}{D_{*}^{2}}\right)\,\left(\frac{2\,k_{B}T_{*}}{3\,h{\nu}}\right)^{8/3}\left(\frac{h{\nu}^{3}}{c^{2}}\right)\,{\int}_{x_{in}}^{x_{out}}\frac{x^{5/3}}{e^{x}\;-\;1}\,dx
\end{equation}

The comparison of  the predictions from equation (3)  with published observations for the ring emission  is shown in Figure 1. To fit the near infrared data, we assume that the innermost
temperature of the ring is 1200 K, appropriate for the sublimation of refractories. From equation (1) and the assumed stellar parameters, this implies that  $R_{in}$ = 1.0 ${\times}$ 10$^{10}$ cm or 0.14 R$_{\odot}$.
Also, because the ring is relatively bright compared to the stellar photosphere,
we assume that $i$ = 0$^{\circ}$.  Finally, in view of the scatter in the
data, a  range of values of $R_{out}$ fit the data about equally well with
 0.4 $R_{\odot}$ $<$ $R_{out}$ $<$ 0.9 $R_{\odot}$, near the value of the tidal radius described below and corresponding to an outer temperature between 600 K and 300 K.     The agreement between the model and
the data shown in Figure 1 is notable, especially since the only 
parameter in equation (3) that is unconstrained by the physical model is the ring inclination.    

Since the ring is opaque, its total mass cannot be determined from the infrared excess alone. A minimum mass can be derived from the assumption that the ring is optically thick even at 15 ${\mu}$m, the longest observed wavelength.  The total surface area of the ring, $A$, is ${\sim}$10$^{22}$ cm$^{2}$. Although the  size distribution and composition of the particles in the ring are unknown,  the maximum likely grain opacity, ${\chi}$, at 15 ${\mu}$m, is  1000 cm$^{2}$ g$^{-1}$ (Ossenkopf, Henning \& Mathis 1992).  In this case, the minimum ring mass, given by $A/{\chi}$, is  10$^{18}$ g.  If the particles are 10 cm in diameter, then ${\chi}$ ${\sim}$ 0.05 cm$^{2}$ g$^{-1}$, and the minimum mass
is 2 ${\times}$ 10$^{23}$ g.  For reference,
Saturn's rings have a total mass of 2 ${\times}$ 10$^{22}$ g and an inner
radius of 0.13 R$_{\odot}$ (Esposito 1993).

\section{RING ORIGIN AND EVOLUTION}

We now consider the origin and evolution of the ring.  Following Debes \& Sigurdsson (2002), we picture that
an asteroid's initial relatively circular orbit was drastically altered  after the Asymptotic Giant Branch evolution of the progenitor to G29-38  with the consequence 
that it strayed within the Roche radius of the star and was tidally disrupted.
The  debris from this disruption-event evolved into the opaque ring that we infer to be present, a model which is  similar to that proposed by  Dones (1991) for the formation of Saturn's rings.  

While a complete description of the tidal interaction between an asteroid and a white dwarf is  complex, in the simplest picture,  
 an asteroid of density, ${\rho}_{a}$ near a star of density,
${\rho}_{*}$ and radius $R_{*}$ is disrupted within a region at separation, $R_{tide}$, given by the
expression (Davidsson 1999):
\begin{equation}
R_{tide}\;=\;C_{tide}\left(\frac{{\rho}_{*}}{{\rho}_{a}}\right)^{1/3}\,R_{*}
\end{equation}
where $C_{tide}$ is a numerical constant of order unity which depends upon the orbital parameters of the asteroid, its rotation and composition.  
For an asteroid with ${\rho}_{a}$ = 3 g cm$^{-3}$ around G29-38, $R_{tide}$  ${\sim}$   1.5  $C_{tide}$ R$_{\odot}$, and therefore, the model dust ring described above lies
within the region where an asteroid would be tidally destroyed.    Once the asteroid is disrupted, the debris of dust particles and rocks undergoes   a complex evolution.  Mutual collisions dissipate energy but not angular momentum so  the system evolves towards
a flat ring composed of matter in circular orbits, reminiscent of Saturn's 
rings (Dones 1991).  Another important process is the inward drift of material towards the central star under the action of the Poynting Robertson effect.  The innermost radius of the dust ring is established
by the location where the particles sublimate.         

We can indirectly infer
the minimum mass of the tidally-disrupted asteroid if the ring material ultimately accretes onto the star.      To account for the observed photospheric metallicity, Koester et al. (1997) estimated
an infall rate of interstellar matter onto G29-38 of 5 10$^{-15}$ M$_{\odot}$ yr$^{-1}$, a rate which which is probably not strongly affected by the
star's pulsations (see Corsico et al. 2002).  Since
interstellar matter is hydrogen-rich, the metal-infall rate is ${\sim}$10$^{-16}$ M$_{\odot}$ yr$^{-1}$.  If accretion occurs from
a rocky ring for 20 years, the
total mass of refractory material in the ring must be at least 4 10$^{18}$ g, implying a radius greater than 10 km.   
 The total mass of accreted matter onto the star is probably much greater than 
this value
since it is likely that G29-28 has  had calcium in its photosphere for much longer than 20 years.    Since  there are metals  in the atmospheres of over 10\% of all white dwarfs (Zuckerman \& Reid 1998), it is plausible 
that G29-38 has been actively accreting for 10\% of its cooling age or 4${\times}$ 10$^{7}$ years.  In this case,  the total mass of accreted material is ${\sim}$4 ${\times}$ 10$^{24}$ g, comparable to
the 2 ${\times}$ 10$^{24}$ g in the asteroid belt of the Solar System (Binzel et al. 2000) or, equivalently, about 1\% of the mass of the asteroids around ${\zeta}$ Lep (Chen \& Jura 2001).  While G29-38 may have been bombarded by many asteroids with radii ${\geq}$ 10 km, it is also possible that the ring has arisen from the disruption of a single object with a radius of 1000 km.

Why did the asteroid stray near the white dwarf?  One possibility is
 that the parent-body to the dust ring was initially in a nearly circular  orbit
where it was always substantially  further from G29-38 than 1 AU so it could survive the second red giant branch evolution of the star.  Additionally, there was at least one Jovian-mass planet  orbiting the main sequence progenitor. As described by Debes \& Sigurdsson (2002), when the Asymptotic Giant Branch progenitor to G29-38  lost over half of
its initial main sequence mass,  a previously stable configuration of planetary orbits could become unstable.  
In this new environment, an asteroid's orbit would evolve dramatically and it could acquire a  high eccentricity by interactions with the Jovian-mass planet(s) (Ford, Havlickova \& Rasio 2001).  The ultimate fate of
an asteroid in its new orbit would be (1) to collide with a perturbing planet or another asteroid,
(2) to collide with the central star (or, as we propose above, be tidally disrupted by
the white dwarf) or (3) to be ejected from the system (see Tremaine 1993). The relative probability of each of these fates depends upon
the particular set of planet and asteroid masses and orbits;  according to 
Debes \& Sigurdsson (2002), there are a number of configurations where the asteroid ultimately collides with the central star. The  duration of this unstable phase might last over 10$^{8}$ years since it is sensitive to the initial conditions of the system.      

\section{RELEVANCE TO OTHER WHITE DWARFS}
Both helium-rich (Dupuis et al. 1993) and hydrogen-rich (Zuckerman \& Reid 1998)  white
dwarfs sometimes show metals in their atmospheres, yet only G29-38 is known to have an infrared excess.
Do the stars with metals also happen to possess dusty rings which have not yet been
 detected? Consider the following evidence.
(1)     Most of the  hydrogen-rich white dwarfs with metals  identified by Zuckerman \& Reid (1998)  have effective temperatures less than 9,000 K.  
It can be shown from equation (3) that at 2.2 ${\mu}$m,  the expected  ring emission is only  10\% of the photospheric flux and may not be easily recognized. Since most infrared surveys of white dwarfs have been restricted to the 
J, H and K bands, observations at longer wavelengths may be required
to detect the rings (if they exist) around these white dwarfs.  
(2) Dupuis et al. (1993) summarize the detections of metals in 
helium-rich white dwarfs including 12 stars with $T_{eff}$ $>$ 10,000 K. If these systems had rings similar to G29-38's, they would
have infrared excesses greater than 20\% of the photospheric emission at 2.2 ${\mu}$m, unless
the ring inclination was nearly edge-on.  Most of these stars are so faint (m$_{K}$ $>$ 15 mag) 
that the 2MASS data are not  precise enough to determine with confidence
whether there is a 20\% excess.   (See also Wachter et al. 2003 who describe a search for late-type companions to white dwarfs.)  However, one of these 12 stars, G111-54, has been observed with sufficient
precision and the stellar parameters are well enough known (Bergeron et al. 2001) that if $i$ = 0$^{\circ}$, the estimated upper limit to the ring flux 
at 2.2${\mu}$m is only about 20\% of that predicted by a ring model similar to  G29-38's.  If, however, the innermost grains have a maximum
temperature of 1000 K instead of 1200 K, then there could be an opaque ring around the star since the cooler grains would not emit very much at 2.2 ${\mu}$m. (3) Chary, Becklin \& Zuckerman (1999) observed  12 mainly hydrogen-rich white dwarfs with ISO at 15 ${\mu}$m, and according to equation (3), the ring emission should dominate the total flux.    
  Except for G29-38, the sources  either were undetected or the 
emission is dominated by the photosphere of the star with the typical upper limits
to the ring emission being 0.4 mJy.    Three of the stars (WD 0208+396, WD 
0322-019 and WD 1633+433) are cool enough and also far enough from the Sun that 
 the predicted fluxes from ring emission at 15 ${\mu}$m  for $i$ = 0$^{\circ}$ are ${\sim}$1 mJy, and, within the errors and uncertainties, consistent with the available upper limits.   However, for G1-27 (van Maanen's star), the
predicted flux from the ring for $i$ = 0$^{\circ}$ is near 10 mJy, depending upon the exact inner and outer radii of the ring, which is well in excess of the observed upper 
limit.   Finally,  there is no
detected infrared excess  around the hydrogen-rich white dwarf G238-044 
which has an effective temperature of 20,000 K, so the dwell time for metals is only a few days  (Holberg, Barstow \& Green 1997).  With a  stellar radius of 8.4 ${\times}$ 10$^{8}$ cm (Provencal et al. 1998) and a distance from the Sun of 25 pc (Vauclair et al. 1997), the predicted flux from the ring at 15 ${\mu}$m for $i$ = 0$^{\circ}$ is ${\sim}$5 mJy, depending upon the inner and outer ring radii,  while
the upper limit given by Chary et al. (1999) is 0.2 mJy. In summary, there are probably some  white dwarfs with atmospheric
metals that do not currently possess opaque, dust rings similar to G29-38's. 

Planetary rings in the Solar System are  diverse;  rings around white dwarfs may be as well. There are several possible explanations for the lack of detected infrared excesses around most white dwarfs even if they accrete asteroids.    (1) Portions of Saturn's rings and those of other planets are optically thin (Esposito 1993).  If  particles in a ring are far enough apart and the system is on average optically thin, equation (3) overestimates the infrared flux.   (2)  If   the tidally-disrupted asteroid or comet was composed largely of volatile material, then at the Roche radius,
 solids  would sublimate.
In this case, the ring may be mainly gaseous and emit little infrared radiation.
(3) The accretion may be time dependent.  Asteroids might stray into the Roche region  episodically, and
the ensuing dust ring may be dissipated in a time short compared to the  gravitational settling time out of the atmosphere of the white dwarf which in some cases can be over 10$^{5}$ years. 

G29-38 may have an unusually large
infrared excess because its disrupted asteroid was unusually massive.
For example, in the Solar System, the most massive asteroid, Ceres, carries about half 
of the total mass in the asteroid belt (see Binzel et al. 2000; Michalak 2000). 
There is such a large range in asteroid masses that the rings that result from their destruction 
may vary substantially.
\section{CONCLUSIONS}
1.  We have found that the infrared excess around the white dwarf G29-38 can be explained as an opaque ring of refractory dust with inner and outer radii
of 0.14 R$_{\odot}$ and ${\sim}$1 R$_{\odot}$, respectively.  

2.  We propose that the circumstellar ring, which lies within the Roche radius of the white dwarf, was produced by a tidally-disrupted asteroid.

3.  Accretion from the dust ring can plausibly explain the abundance of calcium in the atmosphere of G29-38.  Either as a bombardment by a series of asteroids or because of one large disruption, the total amount of matter  accreted onto the white dwarf may have been ${\sim}$ 4 ${\times}$ 10$^{24}$ g, about 1\% of the mass of asteroids around ${\zeta}$ Lep.      

  I have had very useful
conversations with E. Becklin, B. Hansen, and B. Zuckerman.   I thank NASA for support.

\newpage
\begin{figure}
\epsscale{1}
\plotone{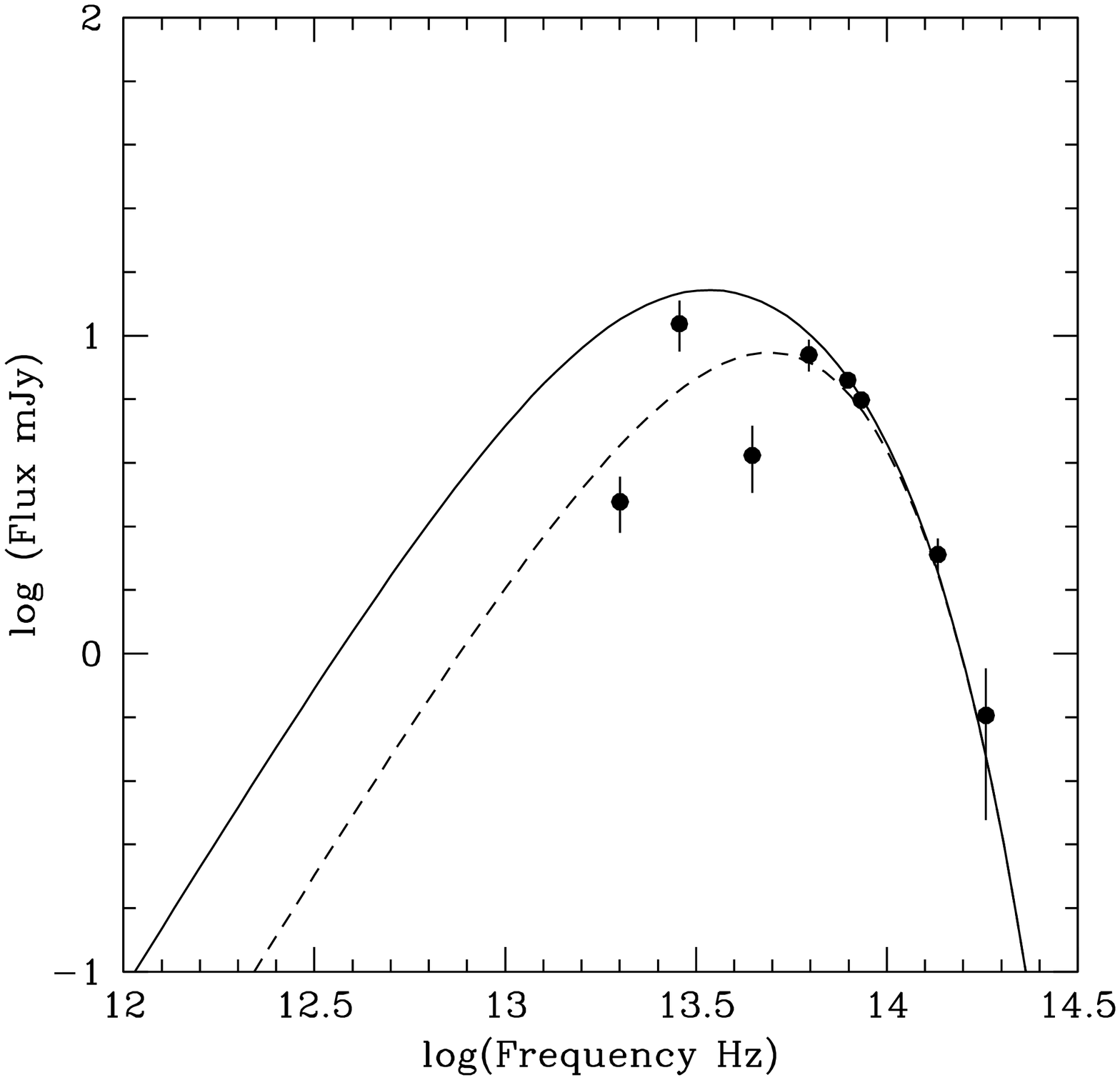}
\caption {A plot of the model spectral energy distributions for the circumstellar dust around G29-38 from equation (3)
compared to observations of the infrared excess from ground-based observations by Tokunaga et al. (1990), Graham et al. (1990) and Telesco, Joy \& Sisk (1990) and  ISO observations by Chary et al. (1999).  At 4.8 ${\mu}$m, we use the revised ground-based flux reported by Chary et al. (1999).    The solid and dashed lines show the 
models where the outermost grains have a temperature of 300 K and 600 K, respectively.  In both cases
the temperature of the hottest grains is 1200 K.}
\end{figure}
\end{document}